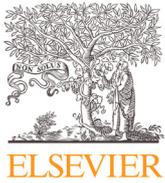
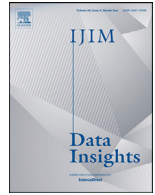

# 2020 U.S. presidential election in swing states: Gender differences in Twitter conversations

Amir Karami [a],*, Spring B. Clark [a], Anderson Mackenzie [a], Dorathea Lee [b], Michael Zhu [c], Hannah R. Boyajieff [d], Bailey Goldschmidt [e]

[a] *School of Information Science, University of South Carolina, Columbia, SC 29208, USA*
[b] *Department of Chemistry and Biochemistry; University of South Carolina, Columbia, SC 29208, USA*
[c] *Department of Psychology, University of South Carolina, Columbia, SC 29208, USA*
[d] *Darla Moore School of Business, University of South Carolina, Columbia, SC 29208, USA*
[e] *College of Nursing, University of South Carolina, Columbia, SC 29208, USA*

A B S T R A C T

Social media is commonly used by the public during election campaigns to express their opinions regarding different issues. Among various social media channels, Twitter provides an efficient platform for researchers and politicians to explore public opinion regarding a wide range of topics such as the economy and foreign policy. Current literature mainly focuses on analyzing the content of tweets without considering the gender of users. This research collects and analyzes a large number of tweets and uses computational, human coding, and statistical analyses to identify topics in more than 300,000 tweets posted during the 2020 U.S. presidential election and to compare female and male users regarding the average weight of the discussed topics. Our findings are based upon a wide range of topics, such as tax, climate change, and the COVID-19 pandemic. Out of the topics, there exists a significant difference between female and male users for more than 70% of topics.

## 1. Introduction

There is a positive correlation between public preferences and actions in politics, indicating that policymakers care about public opinion and public's positions have significant impact on policies (Soroka & Wlezien, 2010). Opinion mining has many application domains such as business, entertainment, politics, marketing (Binali et al., 2009). It was found that gender has a significant impact on opinion and political experiences and gender gaps have political consequences (Bos & Schneider, 2016). For example, the political engagement of women is less than men, female and male citizens have difference political views and activities (Diekman & Schneider, 2010), and women are more likely to vote than men (Feingold, 1994). So, it is important to understand gender gaps because political campaigns could be affected by the gaps. For instance, gender gaps can change strategies of candidates (Diekman & Schneider, 2010).

To inform politicians of public preferences, there is a need to measure public opinion that is "an aggregate of the individual views, attitudes, and beliefs about a particular topic, expressed by a significant proportion of a community" (Davison, 2020). Traditional methods such as interviews, surveys, and observations are essential tools for politicians to track public opinion regarding different issues. However, these methods are labor-intensive, time-consuming, and using a small sample of the target population (Kim & Jeong, 2015).

Social media has become a mainstream channel of communication with a growing popularity across all U.S. populations in the last decade (Karami et al., 2020). In 2021, seven-in-ten Americans use at least one of social media platforms (Auxier & Anderson, 7 Apr, 2021). Social media data can be used to explore public opining toward social events, political movements, company strategies, marketing campaigns, and product preferences (Chen & Zimbra, 2010). Social media has been used by the public during election campaigns to share their opinion regarding different issues (Jungherr, 2016). Among social media, Twitter is a popular platform with more than 68 million U.S. adult users (Newberry, 3 Feb, 2021), including 38.4% female and 61.6% male users (Tankovska, 2021). Twitter offers free Application Programming Interfaces (APIs) to collect data. Politicians and journalists have analyzed Twitter data to look at Twitter data to understand public opinions (Jungherr, 2016). Research using Twitter data is a popular topic and has increased significantly during the last decade (Karami et al., 2020). Researchers have utilized Twitter data for different applications, such as examining happiness, diet, and physical activity (Nguyen et al., 2016), obesity (Ghosh & Guha, 2013), domestic violence (Xue et al., 2019), natural disasters (Ahn et al., 2021), pandemics (Mahdikhani, 2021,

* Correspondence author.
*E-mail address:* karami@sc.edu (A. Karami).

Ridhwan & Hargreaves, 2021), financial markets (Tandon et al., 2021), and sexual harassment (Goel & Sharma, 2020). While there are studies focusing on gender in politics and media, there is no research on comparing normal female and male users regarding the topics of tweets during an election.

## 2. Literature review

Utilizing Twitter for political events has been studied by researchers from different fields and perspectives such as informatics, political science, and communications (Jungherr, 2016) and is one of the up-and-coming research methods, as it is free, accessible, and extremely versatile (Karami et al., 2020). A growing number of studies have investigated how Twitter is used by three groups: political parties, candidates, and normal users. These studies have analyzed different elections in different countries such as the United States (U.S.), United Kingdom, and Germany (Jungherr, 2016). The use of Twitter by the first and the second groups has focused on (1) how parties and candidates use Twitter, such as how the 2012 candidates used Twitter (Adams & McCorkindale, 2013) and (2) the impacts of using Twitter on campaigns, such as exploring the relationship between Twitter discussions and the content of news in the 2012 presidential primary candidates (Conway et al., 2015).

Research on mining the content of tweets has focused on two main directions. The first direction measured the sentiment of tweets regarding candidates (Tumasjan et al., 2011). This direction has utilized both supervised (Bermingham & Smeaton, 2011) and unsupervised (Tumasjan et al., 2011) methods to identify the sentiments of the tweets for each candidate. The second direction utilized text mining methods to understand the semantic of tweets. For example, one study has combined both sentiment analysis and topic modeling to identify positive and negative topics regarding economic issues in the 2012 U.S presidential elections (Karami et al., 2018). A similar study has utilized topic modeling and sentiment analysis to understand public opinion in the 2018 Central Java Gubernatorial Election (Wisnu et al., 2020). Another study proposed a framework based on mining the content of tweets to identify reasons behind the popularity of a politician (Karami & Elkouri, 2019). Topic modeling and co-occurrence retrieval methods were also used to explore topic trends in the 2012 Korean presidential election (Song et al., 2014). Sentiment and semantic analyses were utilized to investigate the predictability of U.S presidential elections (Jahanbakhsh & Moon, 2014) and analyze the behavior of users (Grover et al., 2019).

Current literature shows that there are differences between men and women regarding brain structure complexity (Xin et al., 2019), rational and experiential thinking (Sladek et al., 2010), Latino public opinion (Montoya, 1996), and personality (Feingold, 1994). Regarding political views and activities, it was found that women are more likely to be liberal, vote for Democratic candidates, and participate in political campaigns but have lower levels of political knowledge, awareness, interest, and efficacy (Lizotte, 2020). It was also found that strong differences exist in the way men and women using language on social media (Gordon, 2003). These small-scale qualitative studies mainly focused on the relationship between linguistics elements (e.g., pronoun) (Bamman et al., 2014). Social media has enabled large-scale analyses of the correlation between social comments and demographic variables such as gender.

To understand the direction of public opinion, it is important to obtain a gender-based understanding of public discussions. There exist studies examining the differences between female and male users regarding the content of tweets, such as comparing the social media communication of male and female politicians (Bailey & Nawara, 2019; Beltran et al., 2021; McGregor & Mourão, 2016), female and male reporters (Artwick, 2014), and normal male and female users (Armstrong & Gender, 2011; Chen et al., 2018; Holmberg & Hellsten, 2015; Park et al., 2016). While these studies offer valuable gender-based perspectives to politics and media, there is no research on comparing normal female and male users regarding the topics of tweets during an election.

To address the limitation, this paper uses both quantitative and human coding methods to collect and analyze a large number of tweets posted by female and male Twitter users during the 2020 U.S. presidential election, which is an important political event for U.S. and the rest of the world. The election was held on November 3, 2020, with the highest voter turnout by percentage since 1900. In this election, Joe Biden defeated Donald Trump with more than 81 million votes, the most votes ever cast for a candidate in a U.S. presidential election (Wikipedia 2021). This election faced issues such as the COVID-19 pandemic, racial unrest, and climate change (Wikipedia 2021). Both candidates were very active on Twitter and had millions of followers (Suciu, 2020). For example, Donald Trump posted thousands of tweets during his presidency, had over 88 million followers, and received more than 389 million retweets and 1.6 billion likes by Jan 8, 2021 (Tweetbinder 2021). Considering tweet analysis through a gender lens, this research identifies the topics of tweets and compares female and male users based on the average weight of the topics. This study is based on the following assumption offered by current literature: there is female and male users have different linguistics patterns. We sought to answer the following research questions:

RQ1: What are the topics of tweets posted by U.S. female and male users during the 2020 U.S. presidential election?

RQ2: Was there a significant difference between the female and male users regarding the average weight of each topic identified in RQ1?

To address RQ1 and RQ2, we use topic modeling to identify themes of tweets, human coding to analyze the topics, and statistical analysis to compare female and male users. The paper proceeds as follows. Next, we explain our research methodology, which provides more details on our data collection and analysis. Then, we follow with an explanation of our findings. Finally, we review the study's limitations and discuss future directions.

## 3. Method

This section provides more details on data collection and analysis. We developed a research framework that consists of three phases: data collection and pre-processing, topic modeling and analysis, and statistical comparison (Fig. 1). We applied our framework to tweets posted by female and male users during the 2020 U.S. presidential election.

### 3.1. Data collection and pre-processing

We collected data using a data service provider called BrandWatch. We used queries including "trump" OR "biden" to collect 10,000 tweets per month from June 1, 2020 for each candidate, until the election date, November 3, 2020, for each of 12 swing states defined by the election analytics website FiveThirtyEight (Silver, 2016). We focused on the swing states because these states decide the winner of an election in U.S. (Wikipedia 2021). The swing states include Colorado, Florida, Iowa, Michigan, Minnesota, Nevada, New Hampshire, North Carolina, Ohio, Pennsylvania, Virginia, and Wisconsin.

BrandWatch let us collect tweets posted in the swing states and written in English by female and male individual users (vs. other accounts such as organizations) and exclude retweets. While there are more than two gender identities such as transgender and non-binary, the BrandWatch platform categorizes the gender of users as female, male, and NA (not available). BrandWatch classifies the gender of users based on their first name[1]. The last category means that the platform could not identify the gender. To evaluate the accuracy of the gender assignment process, we randomly selected 1500 tweets and asked two human coders to find the gender of the author of tweets. There were 100% agreement on 1004 authors. We used the 1004 authors to compare the human coding and BrandWatch. We performed a Cohen's $\kappa$ to determine the agreement on gender assignment of users. There was substantial agreement, $\kappa = 0.67$ (McHugh, 2012).

---

[1] https://www.brandwatch.com/blog/product-update-gender-and-account-type-for-twitter/.

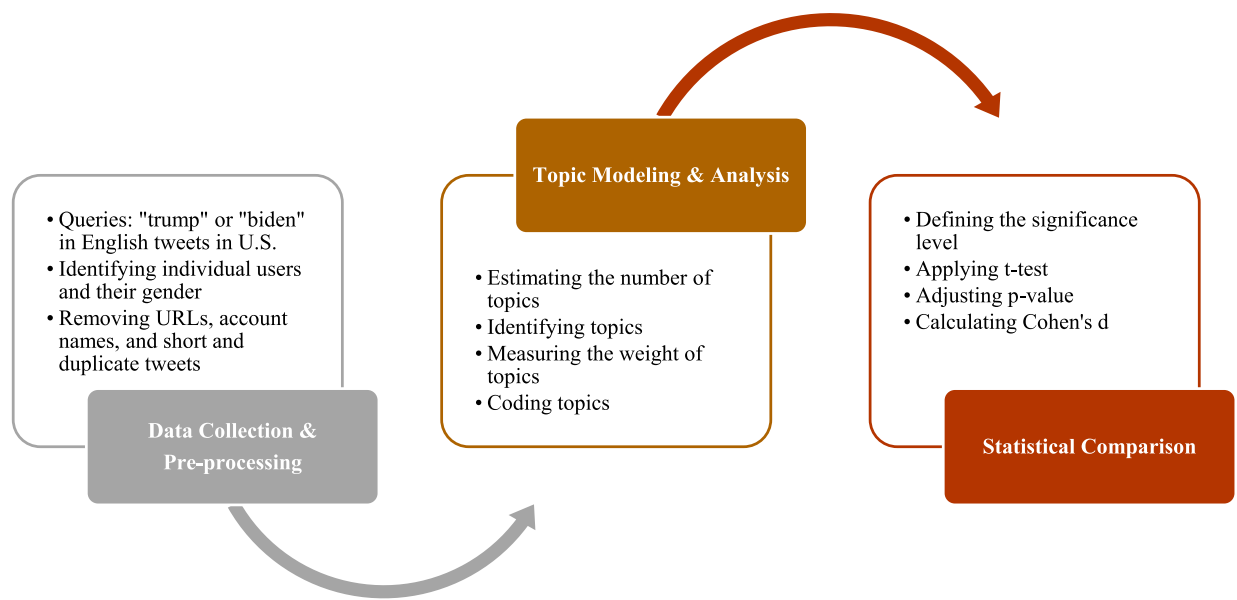

**Fig. 1.** Research framework phases.

In total, we have collected 1,440,000 tweets in total (10,000 tweets × 2 candidates × 6 months × 12 swing states). We have removed verified accounts maintained by popular users such as singers and not being typical user, URLs, and account names starting with @. Then, we filtered out short tweets containing less than five terms and duplicate tweets based on username, date, and state. This process has provided 306,142 tweets, posted by 117,203 (38.3%) female and 188,939 (61.7%) male uses.

*3.2. Topic modeling and analysis*

To address RQ1, we used both text mining and human coding in this phase. Analyzing a large number of documents is a challenging task (Karami, 2019). We utilized topic modeling to identify topics discussed in the tweets. Among different models, Latent Dirichlet Allocation (LDA) is an unsupervised text mining method to disclose hidden themes in corpora (Blei et al., 2003). This popular and valid method has been applied to different corpora in different domains such as health, politics, management, psychology, big data and public opinion mining (Boyd-Graber et al., 2017, Karami et al., 2021, Karami et al., 2021, Karami et al., 2019, Karami et al., 2020, Karami et al., 2021, Kumar et al., 2021, Martinelli, 2022, Mohammadi & Karami, 2022). LDA assumes that each document has a mixture of topics, and each topic includes a distribution of words in a corpus and represents a theme (Blei et al., 2003). For example, LDA assigns "gene," "dna," and "genetic" to a topic representing a genetic theme.

LDA offers two matrices for n documents (tweets), m words, and t topics: probability of each of the words for each topic or $P(W_i|T_k)$ and the probability of each of the topics for each document or $P(T_k|D_j)$. The first matrix shows semantically related words representing a theme, and the second one illustrates the weight of each topic for each document. Before we identify topics and their weight for each tweet, the number of topics should be estimated. To do this, we measured the coherence of topics from 2 to 100 using the C_V method (Rehurek & Sojka, 2010) built in the gensim Python package (Rehurek & Sojka, 2010). It was shown that the C_V method is highly correlated with human ratings (Röder et al., 2015). The optimum number of topics was estimated at 34. Then, we set the Mallet implementation of LDA (McCallum, 2002) at 34 topics and 4000 iterations and removed most common words such as "the". To validate the robustness of LDA, we compared five sets of 4000 iterations and found that the difference between the mean and standard deviation of the log-likelihood of the five sets was not significant.

We qualitatively investigated each topic following two steps to interpret each topic. First, two coders analyzed the top words and top relevant tweets of each topic. The coders answered two questions: (Q1) "Does the topic have a meaningful and relevant theme?" and (Q2) "What is the overall theme of each meaningful and relevant topic?" Q1 helped filter out the topics that do not represent a theme or do not have a relevant them. We used consensus coding (Lim et al., 2015) for Q2 to develop a label for each topic. For example, "trump," "people," "covid," "mask," "care," "wear," "virus," "die," "rally," and "americans" appeared in a topic which coders assigned the label "Trump Administration Response to COVID-19 Pandemic."

*3.3. Statistical comparison*

To address RQ2, we used two-sample t-test built in the R mosaic package (Pruim et al., 2012) to compare female and male users regarding the average weight of topics. For N tweets, we utilized $\frac{0.05}{\sqrt{\frac{N}{100}}}$ (Good IJ. C140 1982) to define the significance level based on sample size (Kim & Ji, 2015). The passing p-value was defined at 0.0009 for 306,142 tweets. We utilized the False Discovery Rate (FDR) method that minimizes both false positives and false negatives (Benjamini & Hochberg, 1995) to adjusted p-values. Other methods such as Bonferroni control false positives, but FDR minimizes not only false positives but also false negatives (Jafari & Ansari-Pour, 2019). Then, we used the absolute effect size using Cohen's d (Sullivan & Feinn, 2012) to measure the magnitude of the difference between female and male users regarding the average weight of topics. The original classes of Cohen's d index (Cohen, 2013) has been extended to six classes, including very small (d=0.01), small (d=0.2), medium (d=0.5), large (d=0.8), very large (d=1.2), and huge (d=2.0) effect sizes (Sawilowsky, 2009). This classification was developed on small sample sizes (Cohen, 2013), and it was found that the average effect size in large samples is less than the one in small samples (Cheung & Slavin, 2016). For instance, one research that measured the effect size for a large sample showed that most effect size values were less than 0.2 (Newman et al., 2008). In addition, the sample sizes used in building the original Cohen's d classification were developed based on 8, 40, 60, 100, 200, 500, and 1000 random samples (Cohen, 2013). To address the restrictions and be in line with the original classification, we obtained 8, 40, 60, 100, 200, 500, and 1000 random tweets and measured the mean of effect sizes of sample sizes.

**Table 1**
Topics of tweets posted in U.S. during the 2020 presidential election.

| Label | Top-10 Words of Each Topic |
| --- | --- |
| Comparing Presidents | trump president obama years history made bush bad job worst remember |
| Wearing Mask in Trump Rally | trump people covid mask care wear virus die rally americans |
| Trump Administration Response to COVID-19 Pandemic | covid trump pandemic virus coronavirus deaths americans response cdc administration |
| Media and Fake News | trump news media lies fake fox truth fact false cnn |
| Trump Campaign, Russia, and FBI Investigation | trump campaign barr crimes fbi russia administration report investigation prison |
| Trump Family | trump donald family wife child father daughter melania son ivanka |
| Vote by Mail | election trump vote mail usps ballot voter office service post |
| Vote for Trump | trump america country president donald american people vote great states |
| Election Analysis | trump election win office lose house republicans senate hold democrats |
| BLM (Black Lives Matter) Movement, Police Violence, and Protesters | trump police violence supporters people cities blm protesters portland riots |
| Trump Foreign Policy | trump war peace deal world middle china iran east israel |
| Trump Supporters | trump supporters signs flags today train front maga car put |
| Supreme Court | trump law court administration supreme rights federal plan judge rules |
| White Supremacy | white trump house racist black proud stand people boys support supremacists |
| Pro-Trump | trump president god love donald family happy bless birthday america |
| Trump Rally | trump rally president campaign live watch great donald make governor |
| Russian Bounty Program | trump putin russian russia troops american military bounties soldiers knew |
| Climate Change | trump head change water point back climate time fire west |
| Anti-Trump | trump shit liar stupid racist idiot dumb moron pathetic corrupt |
| Vaccine and Public Trust | trump fauci health vaccine calls covid trust doctor public medical |
| Mary Trump's Book | trump donald book read story mary president niece report opinion |
| Senate Election | trump vote party republican gop democrat voters candidate election senate |
| Trump Behavior | trump makes sense political support person common behavior lack respect |
| Left Democrats | trump people hate democrats left country fear media evil republicans stop blame |
| Voting For/Against Trump | trump people vote support fuck you supporter hate friends shit |
| Tax | trump money tax pay business paid millions jobs returns rich |
| Trump Town Hall | trump question town answer hall president watch nbc asked tonight |
| Polls | trump vote win state polls election voters red blue electoral |
| Debate | trump joe debate wallace words hear chris time talk night |

## 4. Results

We present our results based on analyzing the 306,142 tweets in two parts. The first part illustrates the topics of tweets. After identifying and analyzing the 34 topics, we found 29 meaningful and related topics. Table 1 shows the topics and their label. The topics cover a wide range of topics from climate change to the COVID-19 pandemic. We observed that "trump" appeared in all topics, but "joe" (biden) only appeared in one topic (debate). This indicates that the center of discussions was Donald Trump.

Most topics were related to U.S. internal issues, including the Trump administration response to the COVID-19 pandemic, tax, vote by mail, BLM (Black Lives Matter) movement, police violence, and protesters, supreme court, white supremacy, vaccine and public trust, left democrats, and senate election. For instance, the third topic focused on the Trump administration response to the COVID-19 pandemic. A relevant tweet claimed the administration "*ignored early coronavirus warning signs. dismissed the seriousness of the threat. attacked the advice of doctors and scientists. failed to institute an adequate national plan for testing and contact tracing. now 150,000 americans have died. this is on trump.*"

Some topics were related directly to Donald Trump and his behavior, family, campaign, supporters, and rallies. For example, the second topic was about the wearing mask policy in place for Trump rallies. Regarding this topic, one user said, "*trump rally will use buses. confined spaces and trump rally attendees will be encouraged to wear a mask but not mandated. rally attendees will be temp scanned but before the buses or after the busses? and buses start at 8:00 am but the rally isn't until 4 pm.*"

Some other topics were about the election, including comparing presidents, election analysis, supporting and opposing a candidate or vote for/against the candidate, polls, and the debate of candidates. For example, the first topic was a discussion on comparing U.S. presidents (e.g., Trump vs. Obama), their performance, and. Regarding this topic, the following tweet compares Obama and Trump: "*a lot of that is obama leftovers. your president made it a point to begin taking credit for obama's accomplishments before he was even in office for 1 year. obama inherited a lousy economy and turned it around. trump inherited a good one, and immediately claimed it as his own.*"

The rest of the topics covered international issues, including climate change, foreign policy, and the Russian bounty program. For example, the author of a tweet related to foreign policy said, "*iran would've had nuke weapons for 5yrs by now if not for nuke deal. iran had uranium for 10 nukes. deal removd nuke mtls, decreasd ability to get uranium. had inspections. trump abandoned our allies & ended deal w/ iran. iran now headed toward nuke weapons. world more dangerous.*" This tweet discussed the nuclear deal with Iran that U.S. was withdrawing from the deal in 2018.

We measured the average weight of topics per tweet. In doing this, we found that polls and foreign policy were the most and the least popular topics, respectively (Fig. 2). The top-3 most popular topics indicate that users were very interested to follow polls, promote their candidate, and talk about how the Trump administration responded to COVID-19. For example, one user compared 2016 poll results and the 2016 elections results in some swing states, "*cnn lv polls nov. 7, 2016 pennsylvania: 47-42 - clinton north carolina: 45-43 - clinton new hampshire: 49-38 - clinton florida 45-45 trump won all of these with the exception of nh, which he lost by less than .5 percent (not 11 points). just sayin.*"

The three least popular topics were Trump's foreign policy, vaccine and public trust, and Mary Trump's book that provided an insider view of the Trump family dynamics (Wikipedia 2021). For instance, one user said, "*this pisses me off! i want to read this book like, now! judge temporarily blocks publication of tell-all book by president trump's niece.*" This tweet referred to the temporary stay of the release of the book ordered on June 30, 2020. However, the order was reversed on July 1, 2020, and the book was published on July 14, 2020 (Wikipedia 2021).

The second part of our findings shows the comparison of female and male users regarding the average weight of the 29 topics discussed in the tweets. Table 2 indicates that there was a significant difference (adjusted p-value <0.0009) between female and male users regarding 21 (73%) topics. Out of the 21 significant differences between female and male users, there were 14 small and 7 very small effect sizes, indicating that the differences were not trivial.

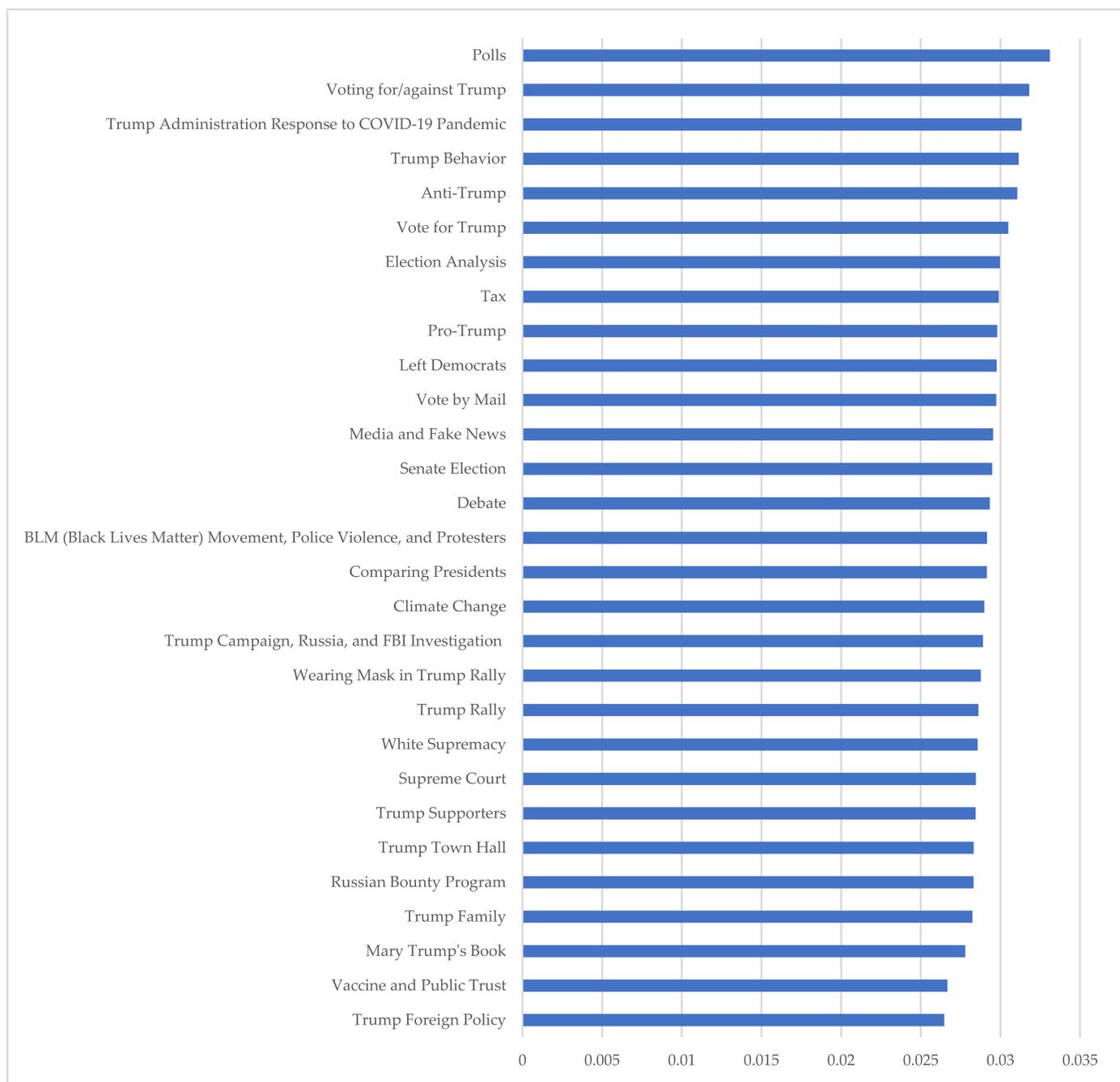

**Fig. 2.** The overall weight of topic per tweet.

There was not a significant difference between female and male users on eight topics: (1) Trump administration response to COVID-19 pandemic, (2) vote by mail, (3) BLM movement, police violence, and protesters, (4) supreme court, (5) anti-Trump, (6) vaccine and public trust, (7) Mary Trump's book, and (8) left democrats. Out of the top-10 topics in Fig. 2, seven topics had a different weight for male and female users (Table 2), including four topics that were discussed more by male users than female users and three topics that were discussed more by female users than male users.

We also identified the top-10 topics based on the average weight of each topic per tweet for female and male users (Table 3). Out of the top-10 topics, there were five common topics between female and male users, including voting for/against Trump, anti-Trump, vote for Trump, Trump administration response to COVID-19 pandemic, and Trump behavior. Pro-Trump, debate, wearing mask in Trump rally, vote by mail, and left democrats were among the top-10 topics of female users, but not male users. On the other side, polls, senate election, election analysis, tax, and comparing presidents were among the top-10 topics of male users, but not female users.

## 5. Discussion

Social media analysis has fewer limitations than traditional methods such as face-to-face interviews. Researchers and politicians commonly use social media data to understand public opinion during elections. It is also important to find differences between the opinion of female

**Table 2**

Statistical comparison of female and male users regarding each topic (F: female users; M: male users; NS: Not Significant; *: adjusted p-value ≤ 0.0009.

| Topics | Adjusted p-value | Results | Cohen's d of Sample Sizes | Effect Size |
|---|---|---|---|---|
| Comparing Presidents | 0.000 | *F<M | 0.1 | Very Small |
| Wearing Mask in Trump Rally | 0.000 | *F>M | 0.2 | Small |
| Trump Administration Response to COVID-19 Pandemic | 0.274 | NS | NS | NS |
| Media and Fake News | 0.000 | *F<M | 0.3 | Small |
| Trump Campaign, Russia, and FBI Investigation | 0.000 | *F<M | 0.2 | Small |
| Trump Family | 0.000 | *F>M | 0.1 | Very Small |
| Vote by Mail | 0.873 | NS | NS | NS |
| Vote for Trump | 0.000 | *F>M | 0.1 | Very Small |
| Election Analysis | 0.000 | *F<M | 0.1 | Very Small |
| BLM (Black Lives Matter) Movement, Police Violence, and Protesters | 0.783 | NS | NS | NS |
| Trump Foreign Policy | 0.000 | *F<M | 0.1 | Very Small |
| Trump Supporters | 0.000 | *F>M | 0.2 | Small |
| Supreme Court | 0.891 | NS | NS | NS |
| White Supremacy | 0.000 | *F<M | 0.2 | Small |
| Pro-Trump | 0.000 | *F>M | 0.2 | Small |
| Trump Rally | 0.000 | *F>M | 0.2 | Small |
| Russian Bounty Program | 0.000 | *F>M | 0.1 | Very Small |
| Climate Change | 0.000 | *F<M | 0.2 | Small |
| Anti-Trump | 0.030 | NS | NS | NS |
| Vaccine and Public Trust | 0.474 | NS | NS | NS |
| Mary Trump's Book | 0.001 | NS | NS | NS |
| Senate Election | 0.000 | *F<M | 0.2 | Small |
| Trump Behavior | 0.000 | *F<M | 0.2 | Small |
| Left Democrats | 0.469 | NS | NS | NS |
| Voting for/against Trump | 0.000 | *F>M | 0.2 | Small |
| Tax | 0.000 | *F<M | 0.1 | Very Small |
| Trump Town Hall | 0.000 | *F>M | 0.2 | Small |
| Polls | 0.000 | *F<M | 0.2 | Small |
| Debate | 0.000 | *F>M | 0.2 | Small |

**Table 3**

Top-10 topics of female and male uses.

| Top-10 Topics of Female Users | Top-10 Topics of Male Users |
|---|---|
| Pro-Trump | Polls |
| Voting for/against Trump | Trump Behavior |
| Anti-Trump | Trump Administration Response to COVID-19 Pandemic |
| Vote for Trump | Voting for/against Trump |
| Trump Administration Response to COVID-19 Pandemic | Anti-Trump |
| Debate | Senate Election |
| Wearing Mask in Trump Rally | Election Analysis |
| Trump Behavior | Tax |
| Vote by Mail | Vote for Trump |
| Left Democrats | Comparing Presidents |

and male users. This paper set out to study topics of tweets posted by female and male users during the 2020 U.S. presidential election. Our results show that female and male users have discussed a wide range of topics in their tweets, and Donald Trump was at the center of many Twitter conversations, which could help to reduce the promotion costs of Trump's campaign. However, the election result shows that being at the center of Twitter conversations may not always be beneficial, as seen from the election results. We also found that there was a significant difference between female and male users regarding the average weight of most topics.

Two surveys developed by the Pew Research Center and Gallup (Brenan, 5 Oct, 2020, Dunn, 2020) have identified 17 important issues for voters, including economy, terrorism and national security, the response to the coronavirus, healthcare, education, race relations, gun policy, violent crime, abortion, immigration, climate change, foreign affairs, taxes, the federal budget deficit, relations with China, relations with Russia, and supreme court. Our analysis shows that Twitter users have discussed eight out of the 17 issues, including race relations, violent crime, tax, supreme court, response to COVID-19 pandemic, foreign policy, climate change, and relations with Russia. Our results do not show that Twitter users did not discuss other issues because LDA identifies major topics, not all topics. Out of the eight issues, response to the COVID-19 pandemic was among top-10 topics of both female and male users, and tax was among the top-10 topics of male users.

### 5.1. Contributions to literature

On a theoretical level, our results increase our understanding of social media communications and how female and male users viewed and engaged with the 2020 U.S. presidential election. This study presents a new way in which female and male users can be analyzed and compared using text mining, human coding, and statistical analysis. Our approach shows how big data analytics can be applied to social media data to provide gender-based insight into political events. Our findings contribute to the literature surrounding topics that are discussed during an election, differences between female and male users, and the most popular topics of female and male users.

The results and polls of the 2020 election show that there was a gender gap (Igielnik et al., 2021, Levitz, 19 Oct, 2020). For example, while 48% men and 55% female voted for Biden, 50% men and 44% women voted for Donald Trump (Igielnik et al., 2021). In line with current literature (Lizotte, 2020), this research confirms that the gender gap existed in the Twitter discussions of swing states in the 2022 election. The gap between men and women was significant in more than 70% of topics. Our results suggest that Policymakers, political parties, and political candidates should consider the gender gaps in their plan While identifying the reasons behind the gap is beyond the goal of this study, prior research investigated why the gender gaps exist in elections (Lizotte, 2020). For example, one reason could be the gender role socialization that leads to different behaviors and attitudes of men and women (Lizotte, 2020). We believe that investigating the origins of gender differences in public opinion can be headed in social media discussions as well.

Most relevant studies developed before 2007 show lower levels of political knowledge, awareness, interest, and efficacy for women (Coffé & Bolzendahl, 2010, Lizotte, 2020). However, 93% of women (vs. 93% of men) adults have access to internet (PewResearchCenter May 2022), 78% of women (vs. 66% of men) adults use at least one social media site (PewResearchCenter May 2022), and 85% of women (vs. 85% of men) adults own a smartphone in 2020 (PewResearchCenter May 2022). In study, this study illustrated that women have discussed a wide range of topics on Twitter during the 2020 election. Considering the facts and our results, there is a need to reconsider the gap between men and women regarding not only political knowledge, awareness, interest, and efficacy but also other issues in elections.

*5.2. Implications for practice*

The proposed approach can be utilized for not only political events but also other non-political events such as health issues (e.g., vaccines). This research also provides some benefits, including offering a cost-effective and time-saving approach to obtain public opinion and evaluate plans and policies, augmenting traditional opinion polls, disclosing the difference between female and male discussions, and illustrating topic priorities in female and male discussions. Our findings can be used for developing social media communication strategies for political campaigns, such as customizing messages for reaching and engaging female and male audiences on Twitter in general.

Popularity of social media platforms and the online nature of the interactions have encouraged of organizations to explore public opinions (Rahmani et al., 2014). Social media discussions have impact on customer perceptions, shopping decisions, sales marketing, and business strategy (Kim & Jeong, 2015). The discussions can be seen as electronic word-of-mouth (eWOM) to convey information from user to user and has impact on new customers and services (Rahmani et al., 2014). Current studies show the application of social media for mining customer feedback (e.g., brand awareness), exploring trends for prediction (e.g., forecast box-office revenue for movies), business decision-making (Rahmani et al., 2014), and reputation comparison (Kim & Jeong, 2015). The approach in this research can be used for product improvement by obtaining information from customers, gaining competitive benefits by developing effective marketing strategies, analyzing public opinion regarding the brand of a company, understanding the concerns of customers, exploring the potential of new products, enhancing customer relationships, and gathering marketplace information.

*5.3. Limitations*

While this research offers a new perspective into understanding the content of social media comments posted by female and male users, there are some notable limitations. First, this paper has considered a binary classification (female or male) of users. However, there are non-binary users outside the binary classification, such as transgender users (Gender 2021). Second, the age of Twitter users is mostly between 18 and 49 years old. The characteristics of this demographic is more likely democrat-aligned, educated, and have high income (Hughes & Wojcik, 2019). Third, the queries are limited to two terms ("trump" or "biden"), which may cause us to miss other relevant tweets. Fourth, our findings should not be generalized to all U.S. female and male users because this study is limited to one social media platform. Fifth, our findings show the importance, not the sentiment, of discussions. Despite these limitations, this study provides novel insights into the topic priorities of female and male Twitter users during the 2020 U.S. election.

## 6. Conclusion

This study identifies the difference between female and male Twitter users regarding the weight of topics in tweets posted during the 2020 U.S. presidential election. We found topics representing a wide range of themes and disclosed that female and male users on Twitter had different priorities of discussions during the 2020 U.S. presidential election. This research highlights that social media can provide gender-based insights during political events. Our work can be used to further understand the differences between social commentary of female and male users on social media. Researchers in information and social sciences could utilize our approach and findings to develop new hypotheses.

There are several future directions for future work to address the limitations of this study, such as including non-binary users, inferring the demographic information of users, and extending the queries. It would also be interesting to analyze non-English tweets, identify and compare social bots, explore the change of topics across different locations and regions, and investigate the reasons behind the gender gap in social media discussions.

**Declaration of Competing Interest**

The authors state that they have no conflict of interest.

**Acknowledgments**

This work is partially supported by the Big Data Health Science Center (BDHSC) at the University of South Carolina. All opinions, findings, conclusions, and recommendations in this paper are those of the authors and do not necessarily reflect the views of the funding agencies.